\documentclass[a4paper,11pt]{report}
\usepackage[utf8]{inputenc}
\usepackage{graphicx, color}
\usepackage{hyperref}
\usepackage{txfonts}
\usepackage{natbib} 
\bibpunct{(}{)}{;}{a}{}{,}

\DeclareRobustCommand{\ion}[2]{%
\relax\ifmmode
\ifx\testbx\f@series
{\mathbf{#1\,\mathsc{#2}}}\else
{\mathrm{#1\,\mathsc{#2}}}\fi
\else\textup{#1\,{\mdseries\textsc{#2}}}%
\fi}

\usepackage{makeidx} 
\makeindex

\begin{document}
\title{\Huge\bf MOSiC\\ an analysis tool for IRIS spectral data}

\author{\normalsize
R. Rezaei\\
%
Instituto de Astrof\'isica de Canarias (IAC), V\'ia Lact\'ea, 38200 La Laguna (Tenerife), Spain\\
\\\vspace{0.3cm} 
\href{mailto:rrezaei@iac.es}{rrezaei@iac.es}, and\,\,\href{mailto:reza5pm@gmail.com}{reza5pm@gmail.com}\\\vspace{0.3cm} 
\href{https://github.com/reza35/mosic}{MOSiC webpage}
}

\maketitle
\tableofcontents

\chapter{\Huge\bf Introduction}
\section{Preface}
The solar chromosphere\index{chromosphere} is a structured layer on top of the photosphere. 
It is optically thin in UV, visible, and near-IR continua, but optically thick 
in strong spectral lines~\citep{carlsson_stein_97, judge_peter_98}.
Although the \ion{Ca}{ii} and \ion{Mg}{ii} lines are the main sources of the radiative loss in the solar chromosphere,
there are many emission lines which form gradually at higher temperatures than the typical low chromospheric value
of 10,000\,K. 
Many of these lines form in the transition region (TR)\index{transition region}
where the temperatures spans a range from
a few $10^4$\,K up to a few $10^5$\,K \citep{gabriel_1976, peter_2001, wilhelm_etal_2007}. 

The  Interface Region Imaging Spectrograph (IRIS)\index{IRIS} is a satellite observatory
launched by NASA on June 27, 2013 \citep{iris_2013}.
It has a 19 cm telescope for UltraViolet (UV) spectroscopy. IRIS has a Sun-synchronous orbit with a period of 97.5 min. 
Spectral bands of IRIS mainly covers the chromosphere and the transition region.

I started to work with \href{http://iris.lmsal.com}{IRIS}\footnote{\href{http://iris.lmsal.com}{http://iris.lmsal.com}}
Mg data a few month after release of the IRIS calibrated data on Oct 2013. 
As I have worked for my PhD thesis and afterward with the {Ca}\,{\sc ii}\,H\index{Ca\,II lines} line observed
with the POLIS instrument~\citep{beck05a} at the German Vacuum Tower Telescope at the Spanish Observatorio del Teide.
it was interesting to work with sharp and clean emission peaks of the Mg\,{\sc ii} lines compared to the one on the 
either sides of the Ca\,{\sc ii}\,H line core. The  Mg\,{\sc ii} lines and nearby continuum are recorded in the
Near UltraViolet (NUV) spectrograph. Beside the NUV spectra, IRIS has also a  Far UltraViolet (FUV) spectrograph. 
The FUV spectra contains many interesting emission lines. 
Some of the FUV lines like the O\,{\sc i}\,135.6\,nm are optically thin \citep{afy}
so the line profiles are like a Gaussian function while other lines like the C\,{\sc ii} line pair
are almost always optically thick and show the self-absorption pattern in the line core. \index{optically thick}
MOSiC fits these lines through a {\em controlled} fitting procedure
in which the degrees of freedom is gradually increased 
and each parameter can span only the specified range.
Measurement errors due to the photon noise (Poisson) and the readout noise (Gaussian)
are considered in the fit \citep{iris_2013}. 
One (two) Gaussians are fitted to optically thin (thick) lines. 
The UV continuum is measured through inspecting the continuum windows in the wing of strong lines (see below). 
%
%
MOSiC uses the \href{http://hesperia.gsfc.nasa.gov/ssw/gen/idl/fitting/mpfit/}{MPFIT}
\footnote{\href{http://www.physics.wisc.edu/~craigm/idl/fitting.html}{http://www.physics.wisc.edu/~craigm/idl/fitting.html}}
package for fitting \citep{mpfit}.\index{MPFIT}

\section{What is MOSiC?}
MOSiC is a collection of IDL programs for profile analysis and Gaussian fitting of the Mg\,{\sc ii}
line along with Gaussian fitting of the C\,{\sc ii}\,133.5\,nm line pair,
the O\,{\sc i}\,135.6, the Cl\,{\sc i}\,135.2, the Si\,{\sc iv}\,139.7 and 140.3 and 
the O\,{\sc iv}\,140.0 and 140.1\,nm lines observed with the IRIS near UV and far UV spectrograph.

\section{The tar file}\index{Tar file}
You can download the whole package or parts of from 
\href{https://github.com/reza35/mosic}{this page}.

After downloading and unpacking the fiels, it creates several directories.  
You should add the path to the some directories (reza\_lib, iris\_reza, other\_lib if you have already installed 
the \href{https://www.lmsal.com/solarsoft/}{SolarSoft}) 
or all directories to your IDL path. 
Other directories includes \href{http://heasarc.gsfc.nasa.gov/fitsio/fitsio.html}{Fitsio programs from Nasa}, 
\href{http://hesperia.gsfc.nasa.gov/ssw/gen/idl/fitting/mpfit/}{MPFIT package}, 
a few routines from IRIS/EIS idl software, as well as a few other individual programs.

\section{Quick overview}
You can get MOSiC working with or without SolarSoft. Therefore, you have to 
de-Rice the level 2 data that you download from internet. Converting the data to level 3 is one of the ways to
convert the Rice-compressed data to a normal Fits data cube. This method is explained in the next section. 
An alternative approach is to use the Python interpreter which reads the Rice-compressed level 2 data correctly
and then write it as a new FITS file. Please note that a few SolarSoft programs which you need for analysis are
included in the tar file but it does not contain bulk of the SolarSoft. 
\break

\noindent MOSiC does not use the header information about the line position.
The user interactively selects lines and the program estimates the amount of the dispersion. Then 
it compares the estimated dispersion with teh one in teh header and when they are fairely close, it takes teh header value and continues. 
If not, the program prompt a warning and stops. 
\noindent Each sub-program checks if the input profile is empty before executing anything, so the program
does not crash in case of a dead pixel/column. 
\noindent Please note that analyzing NUV and FUV data, or every two wavelength band are completely independent.

A common block is used to pass the spectral dispersion to the functions fitting the
profiles. The user does not need to edit any IDL file, except
if you want e.g., to push to fit profiles below the default noise level. 
However the programs are documented so you can read the documentation in each program.

MOSiC was designed to save as much RAM as possible.  
To this end, I rewrote a large number of programs and
configured the programs such that each set of lines should be called
separately. As a consequence, there is one extra step at the beginning where
we reshape the full array, and then read each raster spectra separately. 
MOSiC should be called in the following order:\newline

\begin{tabular}{ll}\index{idl commands}
\rm{$>$ reorder,} & \rm{'/path/to/data/'} \hspace{1.3cm} \\
\rm{$>$ analyze\_iris,} & \rm{'/path/to/data/', /do\_cont} \\
\rm{$>$ analyze\_iris,} & \rm{'/path/to/data/', /do\_mg, /do\_h, /do\_gauss, /do\_improve}   \\
\rm{$>$ analyze\_iris,} & \rm{'/path/to/data/', /do\_o1}\\
\rm{$>$ analyze\_iris,} & \rm{'/path/to/data/', /do\_cl}\\
\rm{$>$ analyze\_iris,} & \rm{'/path/to/data/', /do\_cii, /do\_fast}\\
\rm{$>$ analyze\_iris,} & \rm{'/path/to/data/', /do\_si, /do\_1394}\\
\end{tabular}
\\\\
Please note that /do\_o1 should be executed before any other FUV channel
as we get the systematic FUV velocity form the O\,{\sc i} line.

\subsection{Correction for the orbital velocity}\index{orbital velocity correction}
The temporal gradient due to orbital velocity of the satellite should be corrected.
One way is to use the method described on the IRIS webpage. I use an alternative method and 
calculate the spectral drift separately for NUV and FUV spectra.
In NUV, one of the strong unblended photospheric lines is used after analyzing the NUV continuum data.
In FUV, we first take average profile of each slit position and then use the O\,{\sc i} line center
to construct a temporal curve. 
The velocity maps, e.g. from O\,{\sc i} or Cl\,{\sc i} should not show any 
trace of residual wavelength drift due to orbital velocity of the satellite.

\begin{table}\label{tab:one}
  \caption{MOSiC overview. 
    The quiet command (/do\_quiet) can be used in all cases except
    for the continuum. The plot command (/do\_plot) is useful for evaluation.
    It plots the analyzed profile for each pixel.}\vspace{0.3cm}
  \begin{tabular}{lll}\hline
    wavelength (nm)  &  spectral feature   \\\hline
280  &  photosphere 	          & /do\_cont  \\
279  &  Mg\,{\sc ii} 	          & /do\_mg, /do\_h  \\
279  &  Mg\,{\sc ii} 	          & /do\_mg, /do\_h, /do\_gauss , /do\_improve \\
135.6  &  O\,{\sc i} + C\,{\sc i}     & /do\_o1         \\
135.1  &  Cl\,{\sc i}  	          & /do\_cl         \\
133.5  &  C\,{\sc ii}	          & /do\_cii, /fast \\	
133.5  &  C\,{\sc ii}	          & /do\_cii  \\	
140  &  Si\,{\sc iv} + O\,{\sc iv}  & /do\_si, /do\_1394          \\\hline
\end{tabular}
\end{table}

\subsection{Comments on the Gaussian fits:}\index{Gaussian fit}
A common feature to all analysis programs is that at the very beginning, it analyzes an average profile. 
The user can check the display and make sure that everything is alright. Also it is advisable to run a few
split positions with visualization active so one can make sure that parameters are set correctly. 
The actual run times depends on the
number of spectral points selected for the analysis, the complexity of the observed line profiles,
as well as the performance of the computing facilities. 
Some of the features of the Gaussian fittings are the following:

\begin{description}
  
\item [a)] In all Gaussian fits, each parameter varies in a limited range.
  This (hopefully) prevents the program to return meaningless results.
  Also in several cases, the step size of a given parameter has also a maximum limit, so for instance in case of a line width,
  we know that it spans a limited range of a few pixels, so the maximum step size is about 0.5 pixel.
  
\item [b)] In each case, output of the fit for each pixel is checked
  and the fit is repeated if the result is not acceptable. When the reduced chi-square is still not satisfactory
  a new procedure with random initialization is triggered. This slows down the program but helps in case of noisy data. 
  
\item [c)] The degree of freedom increases gradually, i.e., we always start from a
  single Gaussian fit, then try a double Gaussian fit, ....
  Each fit then provides an 
  initial guess for the next fit. This procedure takes a little bit longer but
  if one starts a fit with 13 free parameter at once, there is a larger chance to have a  failure.
  This holds for all spectral bands except the Cl\,{\sc i} line where we only perform a single Gaussian fit.
  A guided multi-Gaussian fitting method was presented previously by several authors \citep[e.g., ][]{peter_2000, tian_etal_2011}.
  
\item [d)] The quality of the fit is evaluated using the reduced chi-square \citep{bevington_robinson_1992}.
  The reduced chi-square has the chi-square distribution\index{reduced chi-square}
  around unity which is an indicator of a satisfactory fit.

\item [e)] The line width calculated in any program is the 1/e width of the maximum
  which is needed to calculate the non-thermal line width. The non-thermal line width is calculated using the
  {\it iris\_nonthermal.pro} provided through IRIS IDL package. 

\end{description}

\section{Off-limb observations}\index{off-limb observations}
Currently the program runs experimentally on off-limb observations. 
I have not notices a crash but as it was mainly developed for the disk observations, 
it is experimental for the time being. It should be tested against a more diverse set of off-limb data. 
If you use it for off-disk data, please be careful, e.g., by keeping the visualization active during execution for a few slit positions. 
If you notice a bug, please send me an e-mail.

\chapter{\Huge\bf Preparations}
MOSiC reads the data cube slice-by-slice,  you do not need to read the whole data cube at once in any reduction step.
This is particularly helpful for machines with limited memory (RAM).  
As a result, there are a few necessary steps to prepare a level\,2 data cube in a format that MOSiC can use it.
In short, it is converting the level 2 data to a normal FITS file without Rice compression and
then transposing the array to have the wavelength as the first axis. 

\section{Converting level\,2 data to level\,3}\index{level\,3 data}
An easy way to convert level\,2 data to a normal data cube is to convert it to level\,3 data.
There is a program written by Prof.\,Mats Carlsson (iris\_make\_fits\_level3) and is available through \href{http://www.lmsal.com/solarsoft/}{SolarSoft}
and \href{http://iris.lmsal.com/software.html}{IRIS webpage}.
You should download the programs and put them in your IDL path. \newline

\noindent 
$>\,\rm{!path} = \rm{!path} + \rm{`:`+expand\_path('+/scratch/programs/')}$\newline

\noindent This will include all sub-directories in the give path.

\subsection*{For a single file or a few files:}
Now it is very easy to convert a level\,2 data to level\,3:\newline

\noindent$\rm{filepath} = \rm{`/data/obs/iris/spots/mercyry/`}$\\
$\rm{infile} = \rm{[`iris\_l2\_20160509\_131630\_3684510004\_raster\_t000\_r00000.fits`]}$\\
$\rm{iris\_make\_fits\_level3, ww+infile,/all}, \rm{wdir=filepath}$.\newline

If there are more than one level\,2 file, list them in the infile variable.
Please note that size of the output file is about twice as big as the input file.
For very large files, the level\,3 file will be so large that you cannot allocate
enough memory in the IDL. In this case, you can convert different spectral band to level\,3 one by one. 
This is the alternative way compared to using /all command which packs all
level\,3 spectral bands in one FITS file.\newline

\noindent$\rm{iris\_make\_fits\_level3}, \rm{filepath+infile, 0, wdir=ww}$\\
$\rm{iris\_make\_fits\_level3}, \rm{filepath\,+\,infile, 1, wdir=ww}$\\
...

\subsection*{If there are many level\,2 files:}
When there are too many level\,2 files, then it is easier to
proceed as following:\newline

\noindent$\rm{filepath = `/data/obs/iris/mercury/`}$\\
$\rm{files\_d = read\_dir(filepath, filter=`iris\_l2*.fits`)}$\\
$\rm{num = n\_elements(files\_d.files)}$\\
$\rm{save,  filename = `temp.sav`, files\_d}$\\

now restart IDL and activate the path:\newline

\noindent$>\,\rm{idl}$\\
$>\,\rm{!path} = \rm{!path} + \rm{`:`+expand\_path('+/scratch/programs/')}$\newline
$\rm{restore, `temp.sav`}$\\
$\rm{infile= files\_d.files}$\\
$\rm{iris\_make\_fits\_level3, filapath\,+\,infile, /all, wdir=filepath}$\\

\section{Timeseries}\index{Timeseries}
After converging timeserie data to level\,3, it has a 4D structure.
To get MOSiC working, we should converting it back to a 3D form.
One way is to flatten the fourth axis:\newline

\noindent$>\,\rm{iris\_timeseri, `/path/to/data/file/`}$\\

\noindent It converts the 4D data structure to a 3D cube like a normal scan.
Alternatively, you can cut each slit position and
analyze it independently.

\section{FITS manipulation using Python}\index{Python}
The \href{http://www.astropy.org/}{Astropy} package has a powerful FITS interpreter.
It reads the Rice-compressed level\,2 FITS files without any extra cost. 
In the python environment:\newline

\noindent $>>>\,\,\rm{from\,\,astropy.io\,\,import\,\,fits}$\\
$>>>\,\,\rm{hdulist\,\,=\,\,fits.open(`/data/iris\_l2\_20160506\_151211\_3610014160\_raster\_t000\_r00000.fits`)}$\\
$>>>\,\,\rm{hdulist.info()}$\\

\noindent Filename: /data/iris\_l2\_20160506\_151211\_3610014160\_raster\_t000\_r00000.fits\\

\begin{tabular}{llllll}
No. &   Name   &      Type     & Cards &  Dimensions &  Format\\
0   & PRIMARY  &   PrimaryHDU  &   346 &  ()         &     \\
1   &          &   ImageHDU    &    33 &  (669, 1093, 64) &  int16 (rescales to float32)\\   
2   &          &   ImageHDU    &    33 &  (825, 1093, 64) &  int16 (rescales to float32)\\   
3   &          &   ImageHDU    &    33 &  (387, 1093, 64) &  int16 (rescales to float32)\\   
4   &          &   ImageHDU    &    33 &  (674, 1093, 64) &  int16 (rescales to float32)\\   
5   &          &   ImageHDU    &    33 &  (120, 1093, 64) &  int16 (rescales to float32)\\   
6   &          &   ImageHDU    &    33 &  (153, 1093, 64) &  int16 (rescales to float32)\\   
7   &          &   ImageHDU    &    33 &  (762, 1093, 64) &  int16 (rescales to float32)\\   
8   &          &   ImageHDU    &    54 &  (47, 64)  &   float64   \\
9   &          &   TableHDU    &    53 &  64R x 7C  &   [A10, A10, A3, A10, A3, A66, A66]\\   
\end{tabular}\vspace{0.5cm}  
$>>>\,\,\rm{prihdr = hdulist[0].header}$\\
$>>>\,\,\rm{print\,\,repr(prihdr)}$\\
$>>>\,\,\rm{data\,\,=\,\,hdulist[1].data}$\\
$>>>\,\,\rm{data.shape}$\\
$\rm{(64, 1093, 669)}$\\

\noindent The last command actually read the data.
You can inspect the data by plotting some line profiles or showing 2D rasters. 
Now we have the data and we can write it using astropy\footnote{ 
\href{http://docs.astropy.org/en/stable/getting_started.html}{Check the astropy instructions for writing new FITS file}.}

\section{Reorder the data cube}\index{reorder the data cube}
Regardless if you have used Astrpy or SolarSoft, at this point you have a normal FITS file (without compression). 
MOSiC assumes that the data has the traditional format of raster spectra:
the first axis should be the wavelength, the second the slit, and the third the scanning direction.
To do so, the 3D data cube should be transposed, but we do it slightly differently just to make sure that we
do not read the whole cube at once:\newline

\noindent$>\,\rm{reorder, `/path/to/data/file/`}$\\

\noindent Now, we have the final calibrated data and we are ready to start the analysis routines.

\chapter{\Huge\bf The main program}
The core program is called "analyze\_iris.pro".
This program calls a few subroutines and performs the analysis. 
The user does not directly interact with other subroutines during the data analysis. 
I reduce the data in the following order: at first the continuum channel, then the O\,{\sc i}\,135.6\,nm channel,
and then anything else. This is because the 280\,nm continuum and the O\,{\sc i} data are used to estimate the
orbital velocity of the satellite in the NUV and FUV, respectively. 

\section{280\,nm continuum : /do\_cont}\index{analyzing the 280\,nm continuum}
The continuum analysis must be done before any other step.
The program first creates an average spectra and an average intensity along the slit.
The later is useful to select only a restricted range for analysis. 
The user then selects the spectral range in the NUV and FUV (by mouse click).
Please read the comments in the terminal carefully.
The program asks for all major lines observed by IRIS (Table~1). 
It can happen that the program asks you to click around a particular line like the Cl\,{\sc i} which was not recorded.
In this case, you can simply click anywhere as you will not reduce that particular line later.
This selection is done once and then used in the other subroutines. 
The measured spectral dispersion will be printed in the terminal and compared to
the official dispersion in the header. All relevant information are saved. 
The continuum reduction program also creates a quasi-continuum maps in one Jpeg file
(by averaging some spectral bands). 
It is useful for a first look in the data cube, although one can use the IRIS Slitjaw images for an overview.

After that, the 280\,nm range will be analyzed in detail: the program
calculates line parameters for two photospheric lines selected by the user.
I recommend to use a line which is deep, isolated, and not at the border of the spectral range.
Next it computes a photospheric continuum map by analyzing the whole 280\,nm spectra, 
and calculates the photospheric velocities.
The orbital velocity correction happens in this step:
the user answers a few questions and interactively fit a curve to the average velocity as a function of time. 
User enters an order for the polynomial fit.
If you do not get a satisfactory fit by any polynomial, you can enter a negative number: it stops the program.
You can then {\it manually} construct the proper curve, call it $curve$ in IDL and then continue.
The estimated orbital velocity is then applied to the photospheric velocity maps so the two velocity maps
shown at the end of this step must have meaningful rad/blue structure e.g., for granulation or Evershed flow. 
If not, please re-run this step and select a different spectral line.
Also note that a dummy save file is created for the NUV orbital velocity correction.
This is because in some data sets the O\,{\sc i} line is missing.

\section{The O\,{\sc i}\,135.6\,nm : /do\_o1}	\index{analyzing the O\,{\sc i}\,135.6\,nm}	 
				 
The O\,{\sc i}\,135.6\,nm  and C\,{\sc i} line pair at  135.4288/135.584\,nm are nearly always optically thin lines forming
in low chromosphere. In the quiet Sun, the  O\,{\sc i} is stronger than the C\,{\sc i} while the opposite happens
e.g., in flares. 
\begin{figure}
\centering
 \resizebox{13cm}{!}{
  \includegraphics*{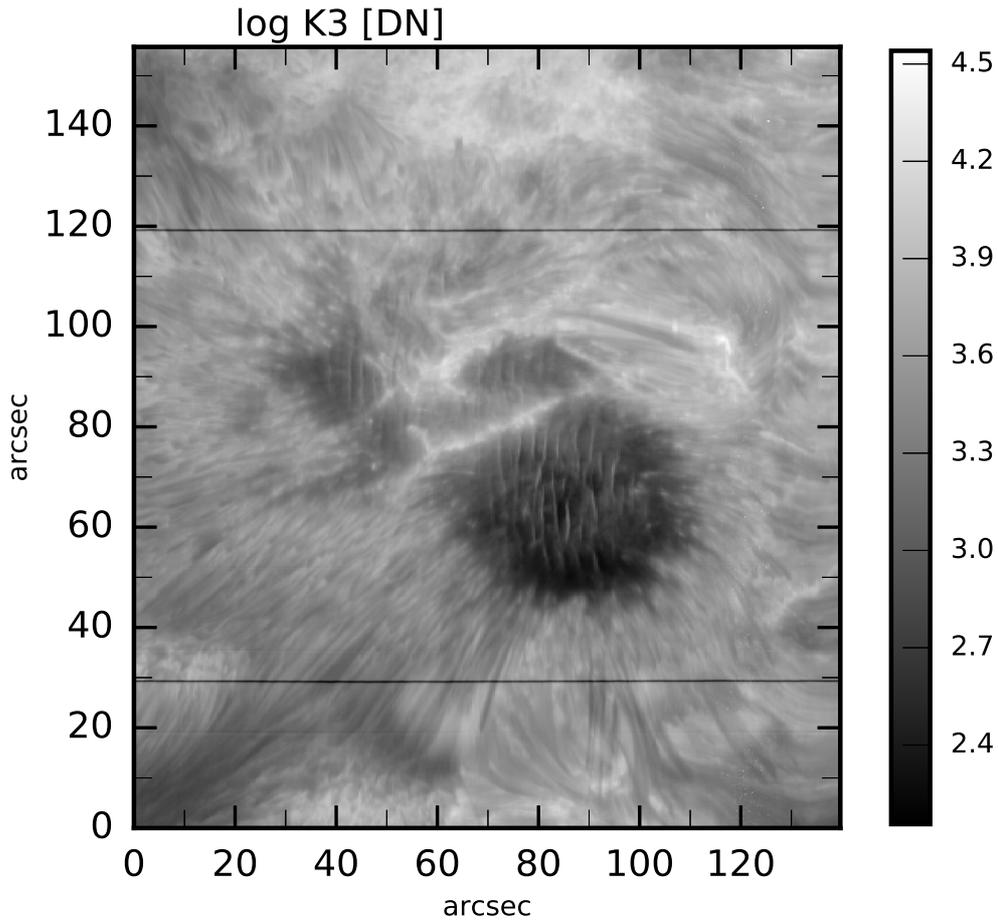}}
  
\caption{Logarithmic map of the Mg\,{\sc ii}\,k$_3$ intensity for large sunspot on 2014/10/24.}
\label{fig:k3}
\end{figure}

At first the program guess the line positions and fits the average profile of the map.
This step is done for all spectral bands.
The program then continues to measure the orbital velocity.
To this end, the program evaluates average spectrums along each slit and constructs the orbital velocity curve.
The user has to answer identical questions as in case of the continuum channel.
The information is saved in a separate file and is used for all other FUV channels.

A multi-line and multi-Gaussian fitting is used to fit these three lines: 
at first we fit a single Gaussian to a restricted spectral range around the O\,{\sc i}\,135.6\,nm line.
Assuming that the Doppler shifts of these three lines are comparable, the program estimates the locations of other lines
and in the next step fits three Gaussians to these three lines. 
Ratio of different line pairs are used to discriminate between a strong signal and a cosmic ray. 
The maps of the 140\,nm continuum, line position, amplitude, and widths get updated progressively except
if the user does not select the /do$\_$quiet option.
If you have created a reasonable orbital velocity curve in the first step,
the final line-core and center-of-gravity velocity maps should not show trace of a velocity gradient along the scanning direction.

\section{The Mg\,{\sc ii}\,h and k : /do\_mg, /do\_h, /do\_gauss}\index{analyzing the Mg\,{\sc ii}\,h and k lines}	 
The program analyzes the line profile and optionally also fits it with Gaussians.  
An example of the profile analysis results is shown in Fig.~\ref{fig:k3}. 
Its main task is to measure the position, amplitude, and widths of the emission peaks. 
The program at first creates a big plot, guess the line positions,
and marks all the important line position and ranges,
including k$_{\mathrm{1}}$ minimum range, k$_{\mathrm{2v}}$, k$_\mathrm{3}$, k$_{\mathrm{2r}}$. The same is done for the h line. 
In addition, it marks two photospheric lines between the k and h lines.
The user can correct if the shown line positions are not satisfactory. This is often the case
close to limb where the line is slightly broader than on the disk center.
the profile analysis program calculates several parameters for the emission peaks.
For a detailed list of the calculated parameters, either check the output variable or read the header of
analyze\_mg.pro. Please note that the program always analyzes the two lines separately (first k, then h) 
so if you want the Gaussian fit for both h and k lines, you have to activate it when you call the program.
				
\begin{figure}
\centering
 \resizebox{10cm}{!}{
  \includegraphics*{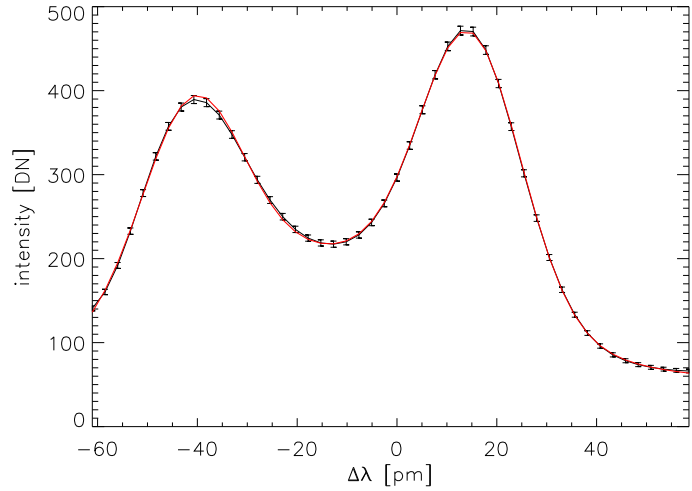}}
  
\caption{Sample Gaussian fits to the Mg\,{\sc ii}\,k line. The Gaussian fit is shown as red curve.
  Errorbars mark the measurement errors due to the photon noise and the readout noise.
  Zero $\Delta\lambda$  marks position of the average profile of the map. }
  \label{fig:mg}
\end{figure}

Before performing a Gaussian fit to the core of the Mg\,{\sc ii} line, the program fits
a parabola to the points outside the k$_{\mathrm{1}}$ (h$_{\mathrm{1}}$) minima for the k (h) line (excluding weak lines). 
The observed profile will be subtracted and then is used for the Gaussian fitting.
Then MOSiC fits a single Gaussian function to the core of the line 
between k$_{\mathrm{1}}$ (h$_{\mathrm{1}}$) minima\footnote{In case of full resolution spectra, this section has 49 spectral pixels.}.
The single Gaussian fit nearly always results in a poor fit except in
sunspot umbra where the self absorption is basically missing.
After that,  the double Gaussian fit is performed using the initial guess from the single Gaussian fit.
The double Gaussian fit has six free parameters: two line centers, amplitudes, and line widths.
Note that we subtracted the base level so we do not need to fit the continuum. 
Finally we perform a triple Gaussian fit with nine free parameters. 
Each of these three fits will trigger a random initialization if the default fit fails. 
Usually a small number of iterations is enough to get a good fit. A sample of Gaussian fit to the Mg\,{\sc ii} line 
is shown in Fig.~\ref{fig:mg}.

\section{The Si\,{\sc iv}\,140.277\,nm : /do\_si}\index{analyzing the Si\,{\sc iv} lines}\index{analyzing the O\,{\sc iv} lines}
Like the case of O\,{\sc i} line, it first plots the average line profiles, and the corresponding Gaussian fits. 
For each profile, the program performs a \underline{five step fitting procedure}: 
The first step is a single-Gaussian fit to the Si\,{\sc iv}\,140.277\,nm line. 
This fit has four free parameters: the continuum, line center, amplitude,and line width. 
A single-Gaussian fit performs a reasonable jobs in some cases but there are many profiles which clearly show a broad tail
component. For this reason, we perform a double-Gaussian fit to the same spectral range.
The  double-Gaussian fit has seven free parameters: the continuum level, two line centers, amplitudes, and line widths.
So far we used a small spectral window around the  Si\,{\sc iv}\,140.277\,nm line. 
In the third step, we fit four spectral lines beside the silicon line. These lines are
the  S\,{\sc iv}\,140.48 line and the three, O\,{\sc iv} lines at 140.00, 140.12, and 140.48\,nm.
This fit has seven degrees of freedom: three free parameters for a single-Gaussian fit to
the Si\,{\sc iv}\,140.277\,nm line, and four amplitudes for the other four spectral lines.
The line widths of the O\,{\sc iv} lines is taken as 1.22 times width of Si\,{\sc iv} line.
The line positions are fixed using offset with respect to the Si\,{\sc iv} line
assuming that all these lines are formed in the same region and having a similar Doppler shift. 
The continuum level measured in the double Gaussian fit is taken as a known parameter in this step.
This penta-Gaussian fit ignores many weak emission lines lines.

The fourth step is another penta-Gaussian fit with ten free parameters.
The main difference compared to the third step is to allow for an independent single Gaussian fit
to the strong O\,{\sc iv}\,140.12\,nm line. At the same time, we fit nine weak spectral lines to improve the residuals
of the fit. The line positions amplitudes, and widths of these lines are all fixed using the other lines.
These nine lines are two O\,{\sc iii}\,140.77 and 140.58\,nm, the  O\,{\sc iv}\,140.79, 
the S\,{\sc iv}\,140.60, the two Fe\,{\sc ii} lines at  140.561 and 139.997, the S\,{\sc i}\,140.126\,nm,
the Ni\,{\sc ii}\,139.903\,nm, and an unknown line at 139.878\,nm for which we use the Ni\,{\sc ii} line width.

\begin{figure}
  \centering
 \resizebox{10cm}{!}{
    \includegraphics*{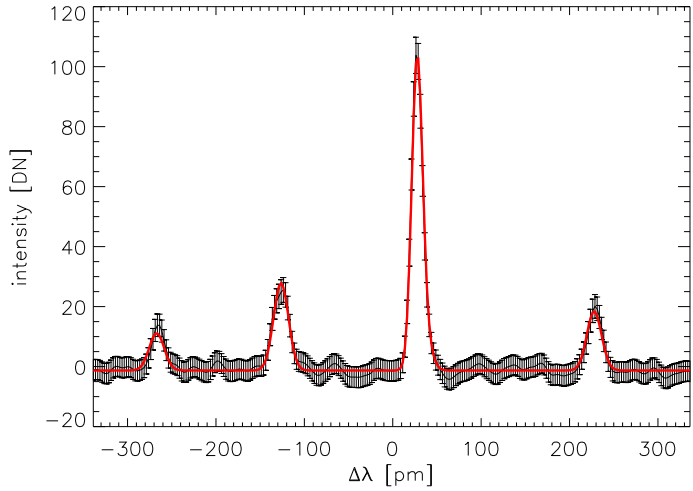}}
 \caption{Sample Gaussian fits to the Si\,{\sc iv} and O\,{\sc iv} lines at 140\,nm and the Gaussian fit (red).
   Errorbars mark the measurement errors due to photon noise and the readout noise. Note the negative continuum level.}
   \label{fig:si}
\end{figure}

The fifth step improves the residuals by fitting a double Gaussian function
to the strong Si\,{\sc iv}\,140.277\,nm line
while treating all other lines similar to the fourth step.
This hexa-Gaussian fit has a similar reduced chi-square distribution compared to the
second penta-Gaussian fit but the peak of the distribution is located closer to the unity.
A comparison of the reduced chi-square between the
third, fourth, and fifth fits show that in some cases, the improvement is not significant
while in case of strong line profiles, or profiles with clear activity signs,
the last hexa-Gaussian fit is clearly the best. An example of this Gaussain fitting approach is seen in Fig.~\ref{fig:si}.

There is a hardwired threshold in the program to skip profiles with a maximum
amplitude below a certain level. For such weak profiles, only the first two fits are performed.
That means the map of e.g., amplitude of the Si\,{\sc iv}\,140.277\,nm  resulting form the single-Gaussian fit 
is complete while map of the same quantity resulting from later fits  has some blank areas.

For the Si\,{\sc iv}\,139.7\,nm, MOSiC fits a single and a double Gaussian function to the line. 
The amplitude of this line is about a factor two larger than the Si\,{\sc iv}\,140.277\,nm line.
However this line is less useful as it is blended with the Ni\,{\sc ii}\,139.332\,nm line.

\paragraph{The electron density:} \index{electron density}
After analyzing the line profiles for the whole map, MOSiC creates maps of some parameters line the non-thermal line width,
the Doppler shift, and estimates the electron density based on the amplitude ratios of the two O\,{\sc iv} lines
at 139.978 and 140.115\,nm. 
To this end, a calibration curve of the amplitude ratio of the two lines are used which is calculated
using the CHIANTI package under the Local Thermodynamic Equilibrium (LTE) assumption \citep{landi_etal_2013_chianti}. 
As LTE does not necessarily holds in all circumstances in the chromosphere and the transition region, this results should be served as
an order of magnitude estimate.

\paragraph{The UV continuum:} \index{UV continuum}
The UV continuum in the 140\,nm range is calculated using probability distribution function of a set of spectral windows,
and then calculating the median of a subset excluding the outliers. 
It still returns a few warm or hot pixels in the final continuum map.
There is an even/odd pattern between successive columns in some maps. 
It can be clearly identified in FFT of big maps, it has a wavelength of a few pixels, and due to its
low amplitude of about 0.5 count, it is not seen in any other intensity map.
Beside that, there is an offset between the upper and lower parts of the chip if one takes old data with old calibrations. 
The bug was fixed and if you download a data now, it does not show this pattern.
				 
In some datasets, the dark subtraction in the calibration procedure creates a negative UV continuum in the
level\,2 data. As such, MOSiC can return negative UV continuum values. No built-in correction is performed
in MOSiC to correct for this artifact. The user has to apply a correction afterwards.

\paragraph{Doppler shifts:} \index{Doppler shifts}
Unlike photospheric spctral lines which have a well-known convective blueshift pattern and have an
average velocity in quiet Sun areas of about zero, the average TR line profiles forming at different temperatures
show significant Doppler shifts \citep{peter_judge_1999}.
MOSiC produces velocity maps at the end of profile analysis. 
These velocities take into account a constant average Doppler shifts for each spectral line. 
That means, e.g., the average Si\,{\sc iv} lines are shift by 4.8\,km\,s$^{-1}$.
This should be born in mind when comparing velocities from different ions.

\begin{figure}
  \centering
 \resizebox{13cm}{!}{
    \includegraphics*{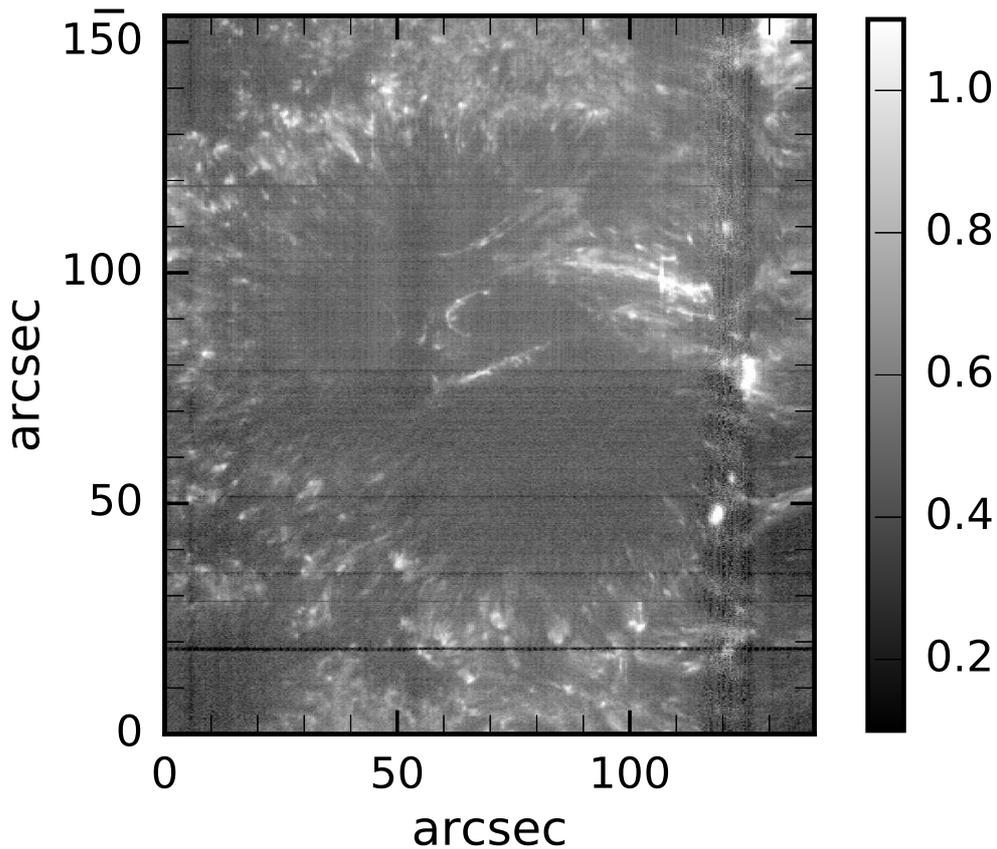}}
 \caption{Map of the UV continuum at 140\,nm for the large sunspot on 2015/10/24 (see Fig.~\ref{fig:k3} for the corresponding Mg\,{\sc ii}\,k$_3$ map)}
   \label{fig:uv}
\end{figure}

\begin{figure}
  \centering
 \resizebox{10cm}{!}{
    \includegraphics*{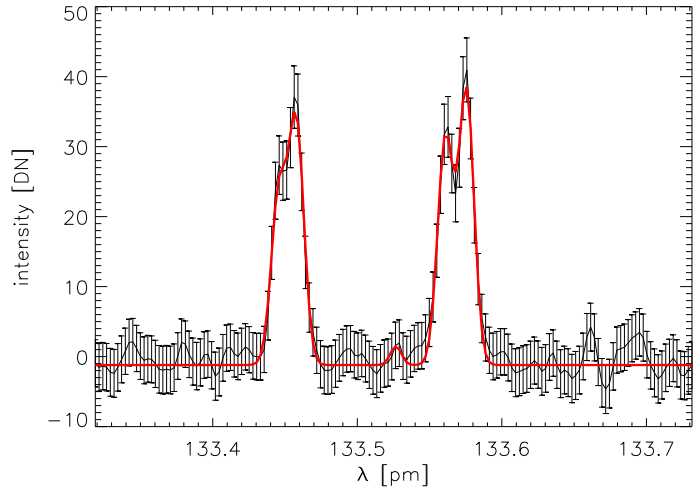}}
 \caption{Sample Gaussian fits to the C\,{\sc ii} line pair (black) and the Gaussian fits (red). 
   Errorbars mark the measurement errors due to photon noise and the readout noise. Note the fit to the weak Ni\,{\sc ii}\,133.52\,nm line.}
   \label{fig:c}
\end{figure}

\section{The C\,{\sc ii}\,133.5\,nm : /do\_cii}\index{analyzing the C\,{\sc ii} line pair}
MOSiC analyzes the pair of C\,{\sc ii}\, lines and the weak Ni\,{\sc ii}\,133.52\,nm line in-between them. 
At first, the program performs single- and double-Gaussian fits
to each C\,{\sc ii}\, line in a multi-Gaussian fitting scheme (together).
The single-Gaussian fit has four free parameters: the continuum, line center, amplitude, and line width.
The double-Gaussian fit has five free parameters as we use the measured continuum level from
the single-Gaussian fit. The five parameters are two line centers, two amplitudes, and a common line width.
This fits both lines but cannot reproduce the central reversal often present in the C\,{\sc ii} lines. 
The quad-Gaussian fit is similar to the double-Gaussian but we consider two Gaussian functions for each of the C\,{\sc ii} lines.
It has ten degrees of freedom, exactly twice  the double Gaussian fit.
The penta-Gaussian fit is basically the quad Gaussian fit
plus a single-Gaussian fit to the Ni\,{\sc ii}\,133.52\,nm line, resulting in 13 free parameters (Fig.~\ref{fig:c}).
Activating /fast will ignore the penta Gaussian fit.

The UV continuum estimated in this step compares well with the one estimated from the Si\,{\sc iv} data.
The line center and center-of-gravity velocities are also calculated.

\section{The Cl\,{\sc i}\,135.17\,nm : /do\_cl}\index{analyzing the Cl\,{\sc i} line}
The Cl\,{\sc i}\,135.17\,nm is an optically thin line like the O\,{\sc i}\,135.6\,nm but is often absent in the recorded spectra. 
MOSiC fits a single-Gaussian function to the Cl\,{\sc i} line profile.
This line provides an alternative for the orbital velocity measurement.
The line-of-sight velocity map from this line compares well with the one from the O\,{\sc i}\,135.6\,nm as both lines
form in the low chromosphere.
Unlike the fit to the Si\,{\sc iv}, C\,{\sc ii} or the O\,{\sc i} lines, this is an isolated spectral line 
so it is more difficult to discriminate a real strong signal from a cosmic ray. 

\section{IRIS coronal lines}\index{cornal lines}
There are two weak coronal lines in the IRIS spectra: the Fe\,{\sc xii}\,134.940\,nm which forms at about
1.5\,MK and the Fe\,{\sc xxi}\,135.408 line which forms at about 10\,MK.
The former is seen in e.g., coronal rain while the later is observed only in hot loops and flares.
These two lines are occasionally observed in 135\,nm range as very broad spectral features,
clearly distinct from the nearby Cl\,{\sc i} or C\,{\sc i} lines.
Beware that they can have a large Doppler shift to either blue or red.
MOSiC in the present version has no option to automatically search for these spectral features.

\section{Evaluating the results}\index{Visualizing results}\index{profile plot}\index{EPS}
There are a few programs included in the package to evaluate the fits.
They plot a histogram of the reduced chi-square and allow the user to move the mouse across a few maps. 
When the user left clicks on a certain pixel, the observed and fit profiles will be plotted on the screen.
With a right click, MOSiC creates the same output as postscript file. 
The maps shown are usually an amplitude map, a map of the reduced chi-square,
and a map of a continuum or wing intensity. \vspace{0.6cm}

\begin{tabular}{ll}\hline
  command &  wavelength range \\\hline
check\_iris\_si1\_fit, `/path/to/save/files/` &  Si\,{\sc iv}\,139\,nm\\
check\_iris\_si\_fit, `/path/to/save/files/`  &  Si\,{\sc iv}\,140\,nm\\
check\_iris\_c\_fit , `/path/to/save/files/`  &  C\,{\sc ii}\,133\,nm\\
check\_iris\_mg\_fit, `/path/to/save/files/`  &  Mg\,{\sc II}\,297\,nm\\
check\_iris\_o\_fit, `/path/to/save/files/`   &  O\,{\sc i}\,135\,nm\\
check\_iris\_cl\_fit, `/path/to/save/files/`  &  Cl\,{\sc i}\,135\,nm\\\hline
\end{tabular}\vspace{0.6cm}

\section{Flaring regions} \index{flares}
In active regions and flaring areas, the C\,{\sc ii} as well as the Si\,{\sc iv} lines can have extreme large amplitudes
and widths \citep[e.g.,][]{peter_etal_2014}.
IRIS FUV detectors show a saturation of the signal at about 1.6$\,\times\,10^4$\,DN, resulting in a top-hat line profile. 
MOSiC was not designed to fit those type of line profiles.
If you see a large value in map of the reduced chi-square, it is worth double checking the observed line profile. 
In particular with long integration times of half a minute, sometimes the line profiles are too complex for the assumed model. 
Also note that in case of flaring areas, some of the blends or weak lines simply disappear.
At the same time, emission lines can have large Doppler shift of the order of several hundred km\,s$^{-1}$. 
In this case, the line can go beyond the range of the normal fitting procedure.
An independent analysis should be performed for such pixels.

\section{Contact}\index{contact}
If you have any suggestion, questions on getting things working, or you have found a bug in the program,
please write an e-mail to \underline{reza5pm$@$gmail.com}. The program will continue to evolve. 
There are a few features I am working on them which will be released in the future versions.

\section*{Acknowledgements}
IRIS is a NASA small explorer mission developed and operated by LMSAL with mission operations executed at NASA Ames 
Research center and major contributions to downlink communications funded by the Norwegian Space Center (NSC, Norway) 
through an ESA PRODEX contract.

\bibliographystyle{aa}
\bibliography{rezabib_nn}


\end{document}